# Quantitative phase imaging via the holomorphic property of complex optical fields

Jeonghun Oh,[1,2] Herve Hugonnet,[1,2] and YongKeun Park[1,2,3]*

[1]*Department of Physics, Korea Advanced Institute of Science and Technology (KAIST), Daejeon 34141, Republic of Korea.*
[2]*KAIST Institute for Health Science and Technology, Korea Advanced Institute of Science and Technology, Daejeon 34141, Republic of Korea.*
[3]*Tomocube Inc., Daejeon 34109, Republic of Korea*
*yk.park@kaist.ac.kr

An optical field is described by the amplitude and phase, and thus has a complex representation described in the complex plane. However, because the only thing we can measure is the amplitude of the complex field on the real axis, it is difficult to identify how the complex field behaves throughout the complex plane. In this study, we interpreted quantitative phase imaging methods via the Hilbert transform in terms of analytic continuation, manifesting the behavior in the whole complex plane. Using Rouché's theorem, we proved the imaging conditions imposed by Kramers–Kronig holographic imaging. The deviation from the Kramers–Kronig holography conditions was examined using computational images and experimental data. We believe that this study provides a clue for holographic imaging using the holomorphic characteristics of a complex optical field.

## I. INTRODUCTION

Holography retrieves the amplitude and phase of a light field, which has been recently exploited for various applications of quantitative phase imaging (QPI). The obtained field is related to the material thickness and refractive index (RI), which enables the imaging of highly transparent objects. Holographic measurements have been utilized in various fields such as rheology [1], nanotechnology [2], biological science [3-6], microfluidics [7-9], and metrology [10,11]. Unfortunately, when an imaging sensor is used, it is only possible to directly measure the intensity of the light fields in the IR, visible, UV, or shorter wavelength regions due to the limited temporal bandwidth of an electronic device. The problem of retrieving the phase from the magnitude information is significantly difficult and historical, known as the *phase problem* [12-14].

With the application of QPI in the biomedical field [15-19], there have been remarkable developments in holographic imaging methods for the reconstruction of complex fields. The retrieval of a complex field falls mainly into two categories: recovering the analytic form of an optical field [20-22] or approaching the correct solution by employing an iterative algorithm [23-25]. There are two approaches for obtaining an analytic solution in terms of methodology: using a reference arm [26,27] or in the non-interferometric regime [28,29]. The complete reconstruction of a complex field without an iterative method not only increases the efficiency of imaging but also eliminates concerns about the possibility of convergence to a correct solution [30]. However, finding an analytical solution with only one intensity image without the assistance of a reference field is challenging. For example, the approach using the transport intensity equation [31,32] provides an elegant way to reconstruct the phase of a complex field by solving a partial differential equation, but requires at least two intensity images and is not capable of imaging all complex optical fields accurately.

Recently, a phase retrieval algorithm using Kramers–Kronig (KK) relations was proposed [33,34]. When a complex field is analytic and square-integrable in the upper half-plane (UHP) of the complex plane, the real and imaginary parts of the field are related through the Hilbert transform of each other, which is called the KK relation. Although attempts to conduct phase retrieval using the properties of a complex analytic field were made extensively in the 1970s and the 1980s [35-38], holographic imaging via the KK relations was applied experimentally only recently [33,39-43].

This study investigates the analyticity of an optical field, focusing on holographic imaging, and describes the significance of analyticity in a 1D phase problem following a procedure similar to that in Ref. [22]. We aim to provide the comprehensive description of holographic imaging exploiting analyticity. The principle of imaging performed using the Hilbert transform is mathematically interpreted using Hilbert microscopy [44] and KK holography [33]. We prove the conditions that make KK holographic imaging hold in terms of a complex analysis. The situation in which the conditions of KK holography are not satisfied was also examined using simulation and experimental data.

## II. HOLOMORPHIC COMPLEX OPTICAL FIELD

For a complex function $f(z)$, it is said to be analytic at $z = z_0$ if the function is complexly differentiable at all points in the neighborhood of $z = z_0$. To avoid confusion with the literal meaning or different usages of the word analytic for real functions, in this study, the analyticity of complex functions is expressed as *holomorphic*; a complex function is holomorphic in a domain or the whole complex plane if the function is analytic at all points in the domain or the complex plane [45].

One can see from the Paley–Wiener theorem that most complex optical functions in imaging systems are holomorphic. The Paley–Wiener theorem [46-49] states that if a function $F(u)$ is supported in $[-C, C]$, expressed as the complex inverse Fourier transform from $\mathbb{R}$ to $\mathbb{C}$,

$$f(z) = \int_{-C}^{C} F(u) e^{iuz} du \qquad (1)$$

is an entire function of the exponential type $C$, which means that it is holomorphic. The fact that such a function is holomorphic can be understood from the Cauchy–Riemann equations in Eq. (1) respects Ref. [45]:

$$\frac{\partial \mathrm{Re}[f(z)]}{\partial x} = \frac{\partial \mathrm{Im}[f(z)]}{\partial y}. \qquad (2)$$

When an objective lens is used in an imaging system, the Fourier spectrum is limited by the numerical aperture (NA) of the lens, as shown in Figs. 1(a) and (b). According to the Paley–Wiener theorem, these complex fields are said to be band-limited and become holomorphic functions. The most complex optical fields in imaging systems have a holomorphic nature. If the 1D regime is considered, the value of $C$ is $2\pi(\mathrm{NA})/\lambda$, where $\lambda$ is wavelength.

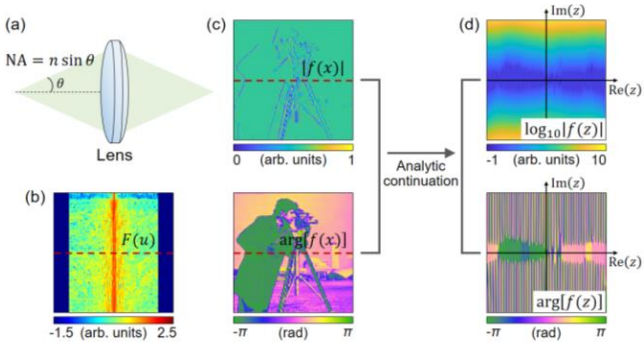

FIG. 1. Holomorphic E field analytic in the whole complex plane. (a) Imaging system with a lens that limits the Fourier spectrum of a sample as NA. $n$ is the RI of a surrounding medium. (b) Amplitude (log scale) of the 1D Fourier transform of a sample field. The Fourier spectrum is supported along with the indicated direction. (c) Amplitude and phase of a sample field. Each line in the direction of the indicated arrow represents the real axis in some complex plane, respectively. (d) Amplitude (log scale) and phase of a complex field expanded to the whole complex plane by analytic continuation. The red dotted lines in (b) and (c) correspond to the function value on a real axis $f(x)$ expanded to the function value in the whole complex plane $f(z)$ in (d) and its 1D Fourier transform $F(u)$.

Equation (1) can be used to obtain a complex function $f(z)$ that matches the value of a real variable function $f(x)$ on the real axis:

$$f(x+iy_0) = \int_{-C}^{C} F(u) e^{iux} e^{-uy_0} du = \mathcal{F}_{1D}^{-1}\left[F(u) e^{-uy_0}\right], \quad (3)$$

with $F(u) = \mathcal{F}_{1D}[f(x)]$, where $\mathcal{F}_{1D}$ represents a 1D Fourier transform. For example, $\sin z$ is the only function whose value on the real axis is the same as $\sin x$, reflecting the identity theorem [50]. This redefinition of a function by extending a domain, such as from $f(x)$ to $f(z)$ [Figs. 1(c), (d)], is called analytic continuation in mathematics [51-53].

### III. COMPLEX ANALYSIS AS A TOOL FOR RETRIEVING PHASE INFORMATION

The KK relations associate the imaginary part with the real part of a complex function $f(z)$ that is holomorphic in the UHP and vanishes at infinity in the UHP [54,55]:

$$\mathrm{Im}[f(x)] = -\frac{1}{\pi} \mathrm{p.v.} \int_{-\infty}^{\infty} \frac{\mathrm{Re}[f(x')]}{x'-x} dx',$$

$$\mathrm{Re}[f(x)] = \frac{1}{\pi} \mathrm{p.v.} \int_{-\infty}^{\infty} \frac{\mathrm{Im}[f(x')]}{x'-x} dx', \qquad (4)$$

where p.v. is the Cauchy principal value. KK relations are often derived using contour integration [56], which is why $f(z)$ vanishes at infinity. The vanishing condition is equivalently expressed using Titchmarsh's theorem [46,57,58] as the Fourier transform of a complex function that needs to be supported on a positive frequency in $[C_{\min}, C_{\max}]$, where $C_{\max} > C_{\min} > 0$. This condition can be understood from the fact that the negative spatial frequencies in Eq. (1) goes to infinity as $\mathrm{Im}(z)$ goes to infinity, thus not reflecting the vanishing condition. This holomorphic property is ensured by the Paley–Wiener theorem. In conclusion, the KK relations can be applied to any band-limited signal with only positive spatial frequencies.

An important application of KK relations is the retrieval of a complex field from its modulus, because many rapidly oscillating quantities in physics, such as light, can only be measured by their modulus. In this case, instead of applying the KK relations to the real part of a complex signal, the real part of the logarithm of a signal can be subject to the KK relations, as follows:

$$g(z) = \log(1+f(z)),$$
$$\mathrm{Re}[g(z)] = \log|1+f(z)|, \qquad (5)$$
$$\mathrm{Im}[g(z)] = \arg(1+f(z)),$$

where $f(z)$ only has positive spatial frequencies. This complex logarithm $g(z)$ is holomorphic if $f(z)$ is holomorphic and $1+f(z)$ does not vanish in the UHP [59]. This proposition can be intuitively understood. $\log f$ at the zeros of $f$ is not only undefined but also non-analytic;

log $f$ is also not square-integrable because log $f$ diverges to infinity at the zeros of $f$.

Whether $1 + f(z)$ vanishes in the UHP can be checked from its value on the real axis using $|f(x)| < 1$, $x \in \mathbb{R}$, and the fact that $f(z)$ is holomorphic and vanishes at infinity. The Rouché's theorem [50,58] bestows a mathematical rationale for the behavior of zeros of a complex function: for two complex functions $f_1(z)$ and $f_2(z)$ holomorphic inside some region $D$ with closed contour $\partial D$, if $|f_1(z)| > |f_2(z)|$ on $\partial D$, then $f_1$ and $f_1 + f_2$ have the same number of zeros inside $D$, where each zero is counted as many times as its multiplicity. If $f_1 = 1$ and $f_2 = f(z)$ and $D$ are set to a large domain in the UHP, including the real axis, Rouché's theorem ensures that $1 + f(z)$ has no zeros in the UHP because $f(z)$ vanishes at infinity in the UHP. Hence, $g(z) = \log[1 + f(z)]$ is holomorphic in UHP if $|f(x)| < 1$ and $x \in \mathbb{R}$.

Finally, $g(z)$ also vanishes at infinity because the Taylor series of $g(z)$ is expressed as a series with no zero-order coefficient, and the powers of $f(z)$ vanish at infinity in a UHP,

$$g(z) = \log(1 + f(z)) = \sum_{n=0}^{\infty} \frac{(-1)^n}{n+1} [f(z)]^{n+1} \quad (6)$$

Note that, from the holomorphic property of $g(z)$, the Taylor series of $g(z)$ converges to $g(z)$ in the UHP.

In summary, the KK relations can be used on the logarithm of a complex function if the function is supported on a closed set of positive frequencies in Fourier space and is of a norm smaller than 1. If there is no zero complex field in the UHP, its logarithm can be reconstructed accurately via the Hilbert transform [37]. The presented proof borrows tools from a complex analysis to derive the conditions of KK holographic imaging. Thus, in the process of retrieving a complex field through KK relations, it can be seen that the growth condition is more decisive than the analytic condition. Furthermore, zeros may be removed to take advantage of phase retrieval using the Hilbert transform. Rouché's theorem can be used to eliminate zero in a UHP.

We can further consider a more general situation, in which a complex function has zeros in the UHP. If $f(z)$ has negative spatial frequencies or $|f(x)| < 1$ is not satisfied for all $x$, $1 + f(z)$ may have zeros in the UHP, so that the KK relations do not hold. In these cases, a holomorphic function of exponential type can be represented by its zeros [38,47]. In particular, such a function is expressed by the Hilbert transform of the logarithm and complex zeros in the UHP [60-62]. When the function satisfies some growth and boundedness conditions, the following equation holds:

$$f(x) = m|f(x)|B(x)\exp(i\mathcal{H}[\log|f(x)|]), \quad (7)$$

where

$$B(x) \equiv \prod_j \frac{z_j - x}{z_j^* - x}, \quad (8)$$

called the Blaschke product [60], $m$ is a constant of modulus one, $\{z_j\}$ denotes the sequence of zeros, and $*$ and $\mathcal{H}$ stands for the complex conjugate and Hilbert transform, respectively. Function values other than the real axis were expanded using an analytic continuation.

It should be emphasized that we can only measure the magnitude of a complex field on the real axis, that is, $|f(x)|$. The sequence of the complex conjugate of zeros, $\{z_j^*\}$, also yields the same magnitude but different complex values of $f(x)$ according to Eq. (8). This ambiguity of zeros is the fundamental origin of the inability to uniquely determine a solution to the 1D phase problem, known as zero flipping [63].

## IV. INTERPRETATION OF QPI METHODS USING HOLOMORPHIC PROPERTIES

We examined two imaging methods, Hilbert microscopy and KK holography, described in terms of holomorphicity and analytic continuation. Consider a beam $S(x)$ scattered from a sample, and a reference beam $R(x) = e^{ikx}$ tilted with respect to the detector plane, representing the off-axis configuration, as shown in Fig. 2(a). Hilbert microscopy [44] utilizes an off-axis configuration. The interference pattern $I$ at the detector is formulated as:

$$I = |R + S|^2 = |R|^2 + |S|^2 + 2\operatorname{Re}(R^*S). \quad (9)$$

In Ref. [44], it is assumed that a sample is a phase object with a uniform amplitude to make the $|S|^2$ constant. Removing the constant components $|R|^2$ and $|S|^2$ leaves $\operatorname{Re}(R^*S)$, where the Fourier transform of $R^*S$ has only positive frequency components under the condition

$$|k| \geq 2\pi(\text{NA})/\lambda. \quad (10)$$

Titchmarsh's theorem ensures that the Hilbert transform of $\operatorname{Re}(R^*S)$ provides $\operatorname{Im}(R^*S)$ and that the sample field $S$ can be reconstructed from $\operatorname{Re}(R^*S)$. The field-retrieval process is shown in Fig. 2(c).

However, unlike the assumption of Ref. [44], even if a sample field has a uniform amplitude, the amplitude of its band-pass-filtered form is not perfectly uniform due to the limited NA of the optical system; thus, $\mathcal{F}[|S|^2]$ is not completely eliminated by the spatial filtering. Only if $|R| \gg |S|$, $|S|^2$ can be fully ignored.

The field retrieval of Hilbert microscopy results from the holomorphic properties of a sample field. In contrast, KK holography uses the holomorphic property of the logarithm of the sample field rather than the sample field itself to improve imaging quality. KK holographic imaging corresponds to the 1D phase problem of retrieving the phase of $R(x) + S(x)$ from the amplitude, $|R(x) + S(x)|$.

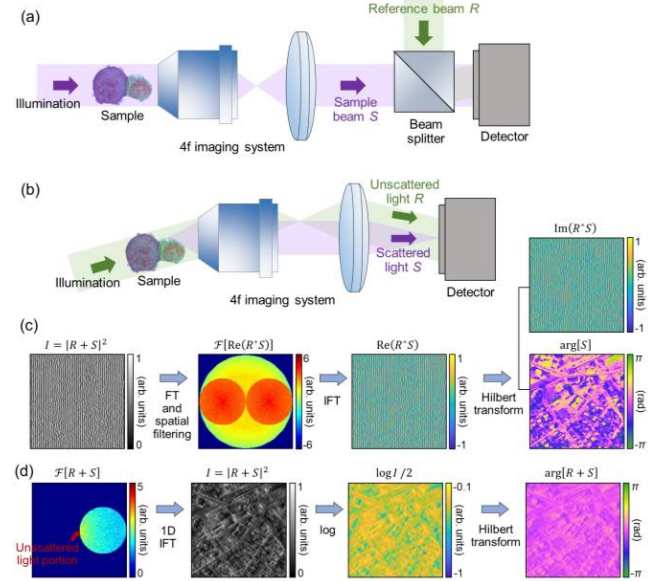

FIG. 2. (a) Schematic diagram of off-axis holography, a setup for Hilbert microscopy. A sample beam is generated by a sample and an illumination, and the interference pattern with a reference beam is acquired from a detector. (b) Schematic diagram of KK holography. The unscattered light from a sample serves as a reference beam. Because $R + S$ is the same as a sample beam itself, no interference fringe appears on the detector. (c) Field retrieval procedure in Hilbert microscopy Spatial filtering of the DC component of an intensity distribution provides the Fourier transform of $2\text{Re}(R^*S)$. The Hilbert transform reconstructs $R^*S$. (d) Process of the field retrieval in KK holography. The Fourier transform of $R + S$ is translated so that the Fourier spectrum does not have negative-frequency components. The intensity $|R+S|^2$ is obtained from a detector. The Hilbert transform of $\log|R+S|$ renders the phase of $R + S$ when satisfying the two conditions; the sample field is reconstructed. FT and IFT are the Fourier transform and the inverse Fourier transform, respectively. Images in the Fourier space are represented by a log scale.

The holographic imaging method using the KK relations confers the Hilbert transform relation to the real and imaginary parts of the logarithm function of $R + S$ with $|R| = 1$,

$$\begin{aligned}\log(R+S) &= \log|R+S| + i \arg(R+S) \\ &= \log|R+S| + i\mathcal{H}[\log|R+S|],\end{aligned} \quad (11)$$

when two assumptions are satisfied: (i) $|R| > |S|$ for all $x$ and (ii) $|k| \geq 2\pi(\text{NA})/\lambda$. These conditions were derived in terms of the complex analysis described in the previous section. Compared with the condition of Hilbert microscopy $|R| \gg |S|$, the condition $|R| > |S|$ is a relaxed requirement.

In the non-interferometric KK holographic imaging [33], the unscattered light takes the role of a reference beam $R$ [Fig. 2(b)]. The samples that can be imaged are limited to weak-scattering objects due to the condition $|R| > |S|$ and the incident angle of the illumination should match the NA of the objective lens, satisfying the condition $|k| \geq 2\pi(\text{NA})/\lambda$. Figure 2(d) shows the process of field reconstruction with the Hilbert transform. The intensity distribution of an optical field, whose negative frequencies are suppressed, is measured in a detector.

Intriguingly, Hilbert microscopy and KK holography require similar imaging conditions. Hilbert microscopy can also be considered in the non-interferometric regime [Fig. 2(b)]. Using the same illumination scheme as in non-interferometric KK holography, the same imaging conditions can be reached when using Hilbert microscopy with $|R| \gg |S|$. Because $|R|$ is larger than $|S|$, the solution given by Hilbert microscopy approaches the field retrieved by KK holography. We simulated Hilbert microscopy and KK holography in this situation [Fig. 2(b)] while satisfying the condition of Eq. (11) and by varying the amplitude of the unscattered light, as shown in Fig. 3. Even when $|R|$ is not significantly larger than $|S|$, Hilbert microscopy shows a high correlation with the solution of KK holography. When $|R| > 4|S|$ [Fig. 3(d)] is used, it can be confirmed that the two solutions are almost identical (correlation of 0.9997). Note that $|R| > a|S|$ implies that $|R| > a|S|$ holds for every pixel of an image, where $a$ is a constant.

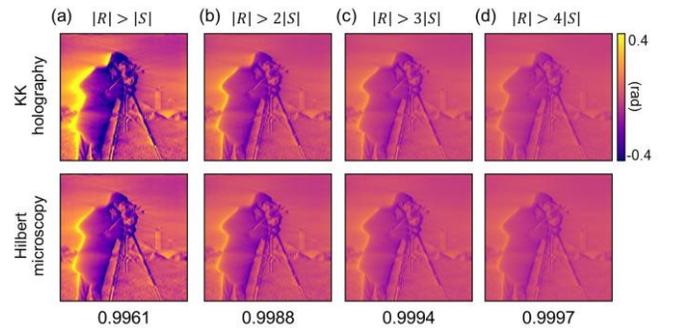

FIG. 3. Comparison of the retrieved phases in KK holography and Hilbert microscopy. As $|R|$ becomes larger, the solution of Hilbert microscopy converges to the complex field reconstructed in KK holography. The number below the reconstructed image is the correlation of the complex fields obtained by the two methods. When $|R|>|S|$, the reconstructed field from KK holography equals the ground truth.

## V. DEVIATION FROM THE CONDITIONS OF KK HOLOGRAPHY

If the two conditions invoked in the previous chapter are satisfied, KK holography provides a correct solution [Fig. 4(a)]. Here, we compare the effects of either of the two conditions being violated. Figures 4(b)–(e) shows the results obtained by changing the amplitude of the unscattered light, while the condition $k \leq -C$ holds, where $C$ is $2\pi(\mathrm{NA})/\lambda$ corresponding to 50 pixels. From the situation in Fig. 4(d), the retrieved phase is significantly different from the ground truth. In particular, it can be seen that the deviation is larger in the part where the phase changes rapidly.

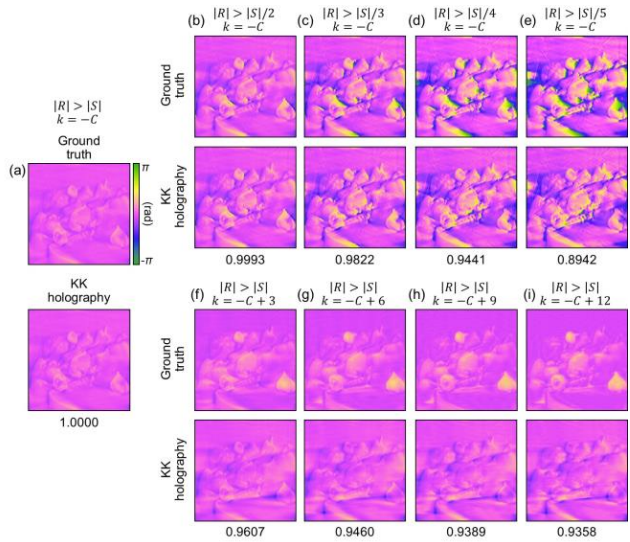

FIG. 4. (a) Reconstructed phase image when the two conditions of KK holography hold. (b)–(i) Reconstructed phase image in the presence of violation for the two conditions with changing (b)–(e) the amplitude of the unscattered light and (f)–(i) the number of deviated pixels from $k = -C$, where $C$ is $2\pi(\mathrm{NA})/\lambda$. The number below the retrieved phase image means the fidelity of the complex field reconstructed by KK holography.

However, while the condition $|R|>|S|$ is maintained, the fidelity is examined by changing the deviation from the condition $k \leq -C$ [Figs. 4(f)–(i)]. $k = -C$ is a situation in which the condition is most tightly satisfied, and the retrieved phase is observed while moving the illumination from a few pixels in the Fourier space. One pixel is the inverse of the image size in real space. As expected, it can be seen that the fidelity decreases as the condition is violated. We also observed that the error of the retrieval given by the deviation is larger for pixels with a high phase value.

We experimentally investigated the effect of deviation from imaging conditions on KK holographic imaging. To clearly inspect the difference in the imaging results of a bead, complex fields obtained from four oblique incident angles are synthesized in Fourier space using a synthetic aperture method [64] to obtain a single-phase image [Fig. 5(a)]. Unless synthetic aperture microscopy is employed, it is difficult to recognize the complete shape of a bead because the spatial frequency distribution of the bead is asymmetrical. We imaged 10-μm-diameter polystyrene beads (RI of 1.5983 at 532 nm, Sigma-Aldrich, 72986-5ML-F) immersed in oil media with different RIs using a conventional off-axis holography setup [65]. Varying the RI of the immersion enables adjustment of the amount of scattering from the sample so that the conditions of KK holography become violated for a low RI of the immersion.

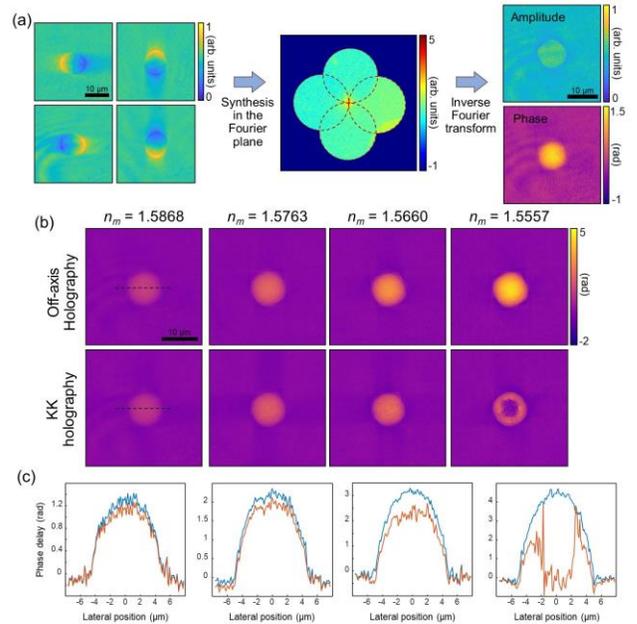

FIG. 5. (a) Reconstruction of a bead image using synthetic aperture microscopy (Left) Amplitude images of a bead acquired at four oblique incident angles. (Middle) The Fourier transform of the four complex fields synthesized in the Fourier space. (Right) Amplitude and phase images of a bead obtained by the inverse Fourier transform. (b) Reconstructed phase of beads in media with different refractive indices using off-axis holography and KK holography. (c) Line profile of the retrieved phase of beads in off-axis holography (blue line) and KK holography (orange line). The dashed line indicates the location where the line profile is obtained. The RI of the medium was obtained at a wavelength of 532 nm.

As the RI difference between the bead and medium increases, the phase delay increases, and the intensity of the unscattered light decreases. Figure 5(b) shows the imaging results according to the intensity of the unscattered light. The ground truth was generated using off-axis holography. When the RI of the medium is 1.5868 and the phase delay of a bead is less than 1.4 rad, KK holography retrieves a phase image similar to the ground truth. In this case, we check that the intensity of the unscattered light is greater than the intensity of the scattered light, except at a few positions. However, the larger the phase delay of the bead, the greater the difference between the ground truth and the bead image reconstructed by KK holography. This deviation can be viewed quantitatively in Fig. 5(c). For the immersion medium with a medium RI of 1.5660, there was already a noticeable deviation from the ground truth. When the phase delay increases, the phase sharply decreases at the center of the bead imaged with KK holography, which resembles a hole. The error is larger at the center of the bead because the phase delay increases closer to the center; thus, the violation of the assumption $|R|>|S|$ becomes severe.

It is possible to think quantitatively about the range of a phase delay where the fidelity of KK holography is maintained. When calculated for phase objects with the condition $|R|>|S|$, the phase range for which KK holography accurately works is up to $\pi/3$. This phase delay corresponds to the situation with a medium RI of 1.5868. Even beyond $\pi/3$, KK holography seems to give somewhat correct phase values, observing the bead in a medium RI of 1.5763. This is because the limited phase delay $\pi/3$ is calculated by considering the normal illumination. In KK holography, the imaging configuration addresses oblique illumination, where the spatial frequencies of the scattered light are concentrated in only one direction. Thus, when exploiting oblique illumination, it is expected that the scattered light can be about two times stronger. In conclusion, the KK holographic imaging provides a value similar to the ground truth for a phase delay of $2\pi/3$, corresponding to the situation with a medium RI of 1.5763.

## VI. DISCUSSION

KK holographic imaging was described by the relationship between a holomorphic field and its zeros. It is important to know the position of the zeros because one can fully determine a holomorphic complex field using its zeros and the Hadamard factorization theorem [47,66]. Interestingly, the problem of eliminating zeros is also associated with a minimum phase function in the field of signals and systems [67,68]. Although this study focuses on using the relation of the Hilbert transform by eliminating zeros, it is also possible to infer the shape of a complex field by directly finding zeros [69,70]. Moreover, because the configuration in which the conditions of the KK holographic imaging are violated is the same as the situation in which zeros exist in the UHP, the deviation can be interpreted with the derived complex zeros [38]. However, in the 1D phase problem, it is necessary to create additional conditions for the zeros because there is ambiguity in zeros when only the magnitude of the complex field is given.

The approach using the Hilbert transform is based on the 1D phase problem because each line of the measured intensity distribution is considered to be the real axis in some complex planes. Approaches via complex analysis for the 2D phase problem are also possible but more complicated because it is necessary to consider a hyperplane in which two complex planes are conjugated [63,71].

Due to its quantitative and label-free imaging capability, QPI has been utilized in various field, with emphasis on the interpretation of QPI data using machine learning approaches [72-77]. However, the complication in constructing interferometric microscopy has hindered wider applications. We envision the present method - QPI via the holomorphic property of complex optical fields via the holomorphic property of complex optical fields would serve as an important basis for comprehensive understanding of phase retrieval problems in holographic imaging as well as expand the applicability of QPI.

*Acknowledgments*.—This work was supported by KAIST UP Program, BK21+ Program, Tomocube, National Research Foundation of Korea (2015R1A3A2066550), and Institute of Information & communications Technology Planning & Evaluation (IITP; 2021-0-00745) grant funded by the Korea government (MSIT).

Crystallography **59**, 1881 (2003).